# ASTRONOMÍA EN LA ESCUELA: SITUACIÓN ACTUAL Y PERSPECTIVAS FUTURAS

María Iglesias, Cynthia Quinteros y Alejandro Gangui

CEFIEC-FCEyN-UBA
Intendente Güiraldes 2620, C1428 EHA, Ciudad Autónoma de Buenos Aires
gangui@df.uba.ar

**RESUMEN:**

Los contenidos mínimos de la educación básica y los programas de estudio de los docentes de ciencias, incluyen varios tópicos en física y astronomía, desde los más simples hasta cubrir temas tales como la fusión nuclear para explicar la evolución de las estrellas o la geometría del espacio-tiempo para una aproximación a la cosmología moderna. En todos esos temas, especialmente en los simples, surgen recurrentemente ideas alternativas, pues ambas poblaciones llegan al aula de ciencias con modelos pre-construidos y consistentes del universo que los rodea. En este trabajo presentamos una serie de interrogantes básicos que, en su conjunto, llevan a reflexionar acerca del estado de situación de la enseñanza-aprendizaje en astronomía en el ámbito de la educación formal a la vez que se plantea la necesidad de dar continuidad, en nuestro país, a investigaciones en el área. En este sentido, se esboza sintéticamente nuestra futura intervención en el campo de la investigación, con la finalidad de contribuir al diagnóstico situacional de los docentes en formación y alumnos en escolaridad primaria, y a partir de la cuál pretendemos desarrollar herramientas didácticas que contribuyan a mejorar su educación formal. Por ultimo, establecemos algunas conclusiones preliminares.

**ABSTRACT:**

Both the basic educational contents for students and study programs for science teachers include several topics in physics and astronomy, from the simplest ones to others as advanced as nuclear fusion to explain stellar evolution and space-time geometry for an approach to modern cosmology. In all these subjects, and most often in the simplest ones, alternative conceptions emerge, as both groups reach science course with preconstructed and consistent models of the universe surrounding them. In this work we present a series of basic questionings that make us reflect on the present situation of the teaching-learning relationship in astronomy within the framework of formal education. We then briefly explain our project aiming at finding the real learning situation of both students and prospective primary-school teachers in astronomical topics and, from the expected results of it, we point towards the need to develop didactic tools that could contribute to improve formal education in astronomy issues.

**Introducción**

Los modelos que construye la ciencia para explicar la realidad parten de las representaciones individuales de los científicos. De igual modo, los niños llegan al aula de ciencias con modelos pre-construidos y consistentes del universo que observan. Los docentes de ciencias en formación también. Estudios realizados en varios países muestran que existe una gran variedad de temas de astronomía en los cuales un alto porcentaje de los futuros docentes de la escuela primaria y media presentan ideas alternativas. La pregunta que cabe hacerse es si están dados los elementos como para que estos docentes entiendan los conceptos básicos de la ciencia (por ejemplo, de astronomía y astrofísica) que luego deberán enseñar. Asimismo, ¿sabemos si existe una adecuada interacción entre los formadores de docentes y los científicos en actividad? ¿Tenemos en claro qué temas esenciales de la materia que los ocupará frente a los alumnos saben correctamente los docentes antes y después de terminada su instrucción? En temas de astronomía, y seguramente en todos los demás, ninguna innovación pedagógica será posible sin antes proveer una adecuada respuesta a estos interrogantes

Para discutir estos temas -que creemos importantes- es necesario ver primero qué sucede en otras latitudes. Debemos luego identificar quiénes presentan problemas con la comprensión de temas científicos o con su transmisión: ¿Solo los alumnos? ¿Quizás también los futuros docentes de ciencias? Los docentes ya formados y con experiencia (y los mismos científicos, con los ya clásicos problemas de transmisión de su ciencia): ¿ellos también? Veremos en lo que sigue que cada uno de estos grupos presenta situaciones donde siempre hay algo por mejorar.

**¿Qué se hace en otros países?**

El reconocimiento de que existen serios problemas de comprensión y de transmisión de temas científicos no es de ayer. La literatura especializada en métodos de enseñanza-aprendizaje, las publicaciones sobre las más comunes nociones alternativas y cómo detectarlas, y los trabajos sobre las diversas maneras de acercarse paulatinamente a un cambio conceptual, son abundantes. Diversas instituciones se han dedicado a tratar de mejorar la cultura científica en las escuelas, para todos sus integrantes, y programas de alfabetización científica están presentes desde hace ya varios años. Algunos ejemplos dignos de mención (sin ánimo de presentar una lista completa) son:

● En EEUU, el programa *Inquiry-Based Science Education*, *Hands-on project*, liderado desde antes de 1990 por el físico norteamericano Leon Lederman;
● En Francia, el programa *La main à la pâte*, impulsado por los físicos Charpak, Léna (astrónomo, en realidad) y Quéré, desde 1996;
● En Chile, el proyecto ECBI: Educación en Ciencias Basado en la Indagación, que funciona por lo menos desde el año 2003;
● En nuestro país, por ejemplo, el Proyecto de Alfabetización Científica del MECyT que, por lo que se ha dado a conocer, está basado en parte en el método francés, y ha sido iniciado hace unos años en un par de provincias tradicionalmente desfavorecidas (en todo sentido) de la Argentina.

Estos programas, por supuesto, no están limitados a la astronomía, sino que abarcan todos los aspectos de las ciencias naturales. Se implementan en las escuelas de educación primaria especialmente (según muchos, la "edad de oro" del aprendizaje) y facilitan el pasaje hacia -y una adecuada articulación con- la escuela media. Los resultados de los equipos de trabajo en los diferentes países han sido reportados en libros, por ejemplo, *Los niños y la ciencia* de Charpak y colaboradores (2006); en sitios web y revistas, por ejemplo, los artículos de Lederman en *Physics*

*Today*; y en jornadas de educación en ciencias, como por ejemplo durante la jornada en la Alianza Francesa, sede de Buenos Aires, del 28 de abril de 2006, en presencia del Ministro de Educación, por mencionar solo algunos ejemplos. Nosotros aquí queremos concentrarnos en los aspectos relacionados con la enseñanza-aprendizaje de la astronomía.

**El planeta Tierra y el Universo**

En el caso particular de la astronomía, existen ciertos temas que se muestran conflictivos a la hora de intentar su cabal comprensión, ya se trate por parte de los alumnos como de los docentes en formación y docentes en ejercicio. Algunos tópicos que repetidamente sobresalen en las publicaciones son:

- Las fases de la Luna: las diferentes fases / iluminaciones de la superficie de nuestro satélite, ¿son debidas a la sombra de la Tierra? Estas fases, ¿surgen como consecuencia de que, desde el punto de vista de la Luna, se produce un eclipse, donde es la Tierra el cuerpo celeste que oculta al Sol?
- Ciclo día-noche: el eterno ciclo de luz y sombra, de días y noches, ¿se debe al movimiento de la Tierra alrededor del Sol? Para los niños más pequeños, ¿el Sol se termina ocultando detrás de las nubes o de las montañas durante la noche?
- Verticalidad en la Tierra y gravitación: el problema de los antípodas, la concepción de la gravitación como atracción hacia el centro del planeta. Los interrogantes: ¿qué significa "caer" hacia abajo?, ¿dónde queda ese "abajo"?
- Las cuatro estaciones: las diferentes estaciones del año, con sus climas y temperaturas característicos, ¿se producen debido a que la Tierra se halla a diferentes distancias del Sol? ¿o es que, para una dada ubicación geográfica, el eje de la Tierra se inclina más en verano que en invierno? (notemos que esto último también se relaciona con las nociones alternativas sobre el ciclo día-noche).
- Composición y forma del sistema solar: las trayectorias planetarias tienen formas pronunciadamente elongadas, se ven como elipses de notable excentricidad (y aquí existe una relación con las diversas nociones sobre las temperaturas en las diferentes estaciones del año). Además: ¿dónde termina nuestro sistema solar?, ¿en la próxima estrella? (Aquí también podemos mencionar la natural confusión generada por la reciente re-clasificación de los planetas y el *affaire* Plutón, relegado a ser -de ahora en más- tan solo un planeta enano…)
- Nuestra ubicación en el Universo: la posición de la Tierra y del sistema solar en el Universo. ¿Ubicación?, ¿respecto de qué?, ¿existe un arriba y un abajo en el Universo?

Estos temas no son los únicos, pero es interesante constatar que son representativos de los contenidos de los programas de estudio de la educación básica para los alumnos, como así también de los currícula de formación docente inicial y de enseñanza media, como veremos más abajo, siempre dentro de la unidad El planeta Tierra y el Universo.

**Problemas para los alumnos: Un par de ejemplos**

Existen varios conceptos sobre la forma de la Tierra y sobre la verticalidad. Estudios realizados hace ya varios años por Nussbaum y colaboradores muestran que los alumnos de la escuela primaria en muchas ocasiones no han incorporado el modelo de un planeta esférico. Su idea de la gravitación es por ello consistente con un cielo y una Tierra planos, y simplemente las cosas caen desde arriba hacia abajo, en cualquier lugar geográfico que se presente. Otros alumnos que sí han incorporado el modelo de una Tierra esférica (por los motivos que sea: porque han visto imágenes espaciales, por la conversación de los mayores, etc.) a veces no han hecho lo propio con la gravitación. Para ellos la gravitación es como un "río" en el cual está "sumergida" la Tierra y por ello alejarse excesivamente de la parte de arriba de la esfera (la parte que recibe de frente el agua de ese río) resulta en la caída fuera del planeta. Ya lo decía Ernesto Sábato en uno de sus libros, cuando se refería a la visión de la Tierra de la ciencia medieval europea: "San Isidoro no admitía siquiera la existencia de habitantes en Libia, por la excesiva inclinación del suelo."

Hay por supuesto varios otros ejemplos de concepciones alternativas donde existe una vertical absoluta, independientemente de la presencia del planeta, o donde dicha vertical no existe sino que depende de la ubicación sobre la superficie de la Tierra (Nussbaum, 1979; Sneider y Ohadi, 1998).

También se ha visto una relación entre los conceptos de verticalidad y gravitación con otras áreas que podríamos pensar "distantes". Tal es el caso de nociones alternativas sobre el cambio de estado y transformaciones, como sucede con la enseñanza de la combustión, o bien en el estudio de la constitución de la materia, cuando se trabaja con una primera aproximación al modelo de partículas de la física moderna. Estudios realizados por Driver y colaboradores muestran que niños de entre 11 y 12 años, ante la pregunta de si un "fósforo" ardería o no en el espacio exterior, responden en un 20% que no, pues en el espacio lejano hay ausencia de gravedad. Estos y otros trabajos llevan a concluir que existe una marcada asociación entre la presencia o ausencia de aire y el concepto de la gravedad (Driver, 1985). Los niños a veces explican la atracción gravitatoria como un efecto del empuje hacia abajo ejercido por el aire sobre los objetos, y por ello, donde no hay aire no debería haber "g" (gravitación).

Se ha visto también que los niños emplean al aire como un medio necesario para la interacción a distancia, lo que según los investigadores lo convierte en una suerte de éter, esa extraordinaria sustancia que inundaba el espacio de los físicos (y no solo de ellos) de hace más de cien años. Los cuestionarios y diversos tests implementados indican que, para los alumnos, sería el aire el que transmitiría tanto las fuerzas electromagnéticas como las gravitatorias. Su ubicación llevaría al aire a estar presente en lugares insospechados (antes de llevarse a cabo los estudios) pues también "entre" las partículas de un gas cualquiera, muchas veces, los niños dicen que podría haber aire.

**¿Qué dicen los programas escolares sobre lo que deben aprender los alumnos?**

En temas de astronomía, los contenidos mínimos publicados por el gobierno de la provincia de Buenos Aires para la escuela primaria y media[1] incluyen:

EGB-1: La Tierra y la Luna en el Sistema Solar. Estrellas, planetas y satélites: diferenciación. Utilización de técnicas para orientarse mediante la observación. Luz y sombra. La sucesión del día y la noche. La observación del cielo y el registro de sus características. Comunicación de la información mediante murales y maquetas. Movimiento del Sol, de la Tierra y de la Luna: hechos observables. Las estaciones. Variables atmosféricas. Obtención de información en textos y otros medios de comunicación.

EGB-2: Movimientos reales y aparentes de los astros. Galaxias y estrellas, viajes espaciales. Orientación y puntos cardinales. Construcción y análisis de modelos del Sistema Solar. Inclinación del eje terrestre. Rotación y traslación de la Tierra. Las estaciones. Los husos horarios. Fases de la Luna, mareas. Eclipses. El ciclo lunar, planetas, estrellas. La sucesión del día y la noche.

---
[1] Organización de contenidos, Ciencias naturales, extraído de sitio web: abc.gov.ar

EGB-3: La Galaxia, características. La Vía Láctea y el Sistema Solar. El Universo: modelos cosmológicos. Escalas de distancias astronómicas. Interrogantes sobre el origen y evolución del Universo (aunque debemos decir que sobre el "origen" aun no hay nada escrito (Gangui, 2005)). El ser humano en el espacio. Generación de energía en las estrellas. Evolución estelar. Modelos históricos del Sistema Solar, modelos cosmológicos y de la evolución estelar.

Por su parte, la Ciudad de Buenos Aires incluye para la primaria[2]:

3º Grado: Cambios y permanencias en el cielo. El Sistema Solar: el Sol, los planetas y otros astros. La Tierra como planeta del Sistema Solar.

4º Grado: no se trabajan contenidos de este bloque.

5º Grado: El cielo visto desde la Tierra. La Luna, satélite de la Tierra. Movimientos aparentes de las estrellas. El Sistema Solar: Movimientos de los planetas.

6º Grado: Magnitudes características. El Universo: las galaxias. Telescopios y satélites artificiales

7º Grado: El Universo: El Sistema Solar. Las estaciones. Las fases de la Luna. Los eclipses.

Como se ve, muchos de los temas que más arriba llamamos conflictivos están presentes en estos contenidos mínimos escolares, y es natural que exista una cierta permanencia de representaciones alternativas en los aspectos más básicos de la astronomía.

Si bien ambos diseños curriculares (el de Provincia y el de la Ciudad de Buenos Aires) incluyen los mismos temas, en Ciudad de Buenos Aires es clara la progresión de contenidos propuesta a lo largo de ambos ciclos, como así también sus alcances. Los alumnos pasan del reconocimiento de regularidades y cambios (duración de días y noches en distintas estaciones, cambios en aspectos de la Luna, etc.) en primer ciclo, para comenzar a formarse a partir de 5º Grado una imagen más estructurada de la Tierra y del Universo, dejando para el último año de la primaria las relaciones más complejas relativas al Sistema Solar. Para el primer ciclo los modos de conocer que predominan son la observación, la descripción y el registro, y se avanza, en el segundo ciclo, en el aprendizaje de los modos de conocer como la búsqueda de información bibliográfica, técnicas de sistematización y organización de la información, entre otros. Por su parte, la organización de contenidos propuesta en Provincia no es transparente sobre los contenidos a trabajar en cada grado y muchas veces, en el ámbito escolar, resulta la superposición de temas o sucede que los maestros no tienen en claro qué tópicos seleccionar en cada uno de ellos, algo que finalmente queda relegado a las editoriales, las que "deciden" con sus libros de texto escolares cuáles son los contenidos importantes a enseñar.

En uno u otro caso queda la pregunta y reflexión de si los maestros y maestras que deben enseñar tales contenidos están capacitados para transmitirlos y qué recursos emplean que favorezcan un aprendizaje significativo. Y, más allá de un diseño curricular que esté más o menos desarrollado, ¿no debería ser el docente quien tomase tales decisiones en función del contexto en el cual está inmerso? El diseño curricular, ¿no es tan sólo un marco de referencia? De ser así, el docente debería dominar los contenidos que están implícitos en las formulaciones más generales que se presentan en dichos diseños curriculares.

## Docentes en formación: ¿problemas en astronomía?

---

[2] Diseño Curricular para la Escuela Primaria. Primer ciclo de la Escuela Primaria / Educación General Básica, 2004 – Bloque Los fenómenos naturales. Diseño Curricular para la Escuela Primaria. Segundo ciclo de la Escuela Primaria / Educación General Básica, 2004 – Bloque La Tierra y el Universo

Cuando los docentes descubren (en carne propia) la necesidad de buscar un cambio conceptual -en el área que sea- les resulta más simple trasponer el mismo problema a sus alumnos. Eso parece natural también en astronomía, pero en nuestro conocimiento esta idea no ha sido puesta a prueba en nuestras escuelas con la frecuencia que sería deseable (Camino, 1995; Frede, 2006).

Estudios llevados a cabo en estudiantes para docentes de escuela primaria en otros países muestran lo siguiente:

- En Inglaterra, sobre un total de 41 futuros docentes, 23 respondieron adecuadamente en temas relacionados con el ciclo día/noche. El resto, mostraron concepciones alternativas o indeterminadas (Parker y Heywood, 1998).
- Entre las concepciones alternativas los investigadores definen la Concepción Alternativa Mayoritaria (CAM). En el caso del ciclo día/noche, en otro estudio llevado a cabo en EEUU, la CAM fue la revolución de la Tierra alrededor del Sol ya mencionada más arriba (Atwood y Atwood, 1995).
- En temas relacionados con las fases de la Luna, solo 6,6% de 76 individuos mostraron un conocimiento científico justo del fenómeno. El resto mostró toda una variedad de concepciones alternativas, entre la que se destaca (CAM) la sombra de la Tierra (Callison y Wright, 1993). Esto corresponde a la así llamada "teoría del eclipse", donde es la Tierra la que oculta la luz solar, como también lo mencionamos más arriba.
- En otro estudio, también relacionado con las fases de la Luna, se pudo comprobar que luego de la instrucción en temas de astronomía, el porcentaje de individuos con concepciones alternativas disminuye fuertemente. Conclusión: los docentes en formación vienen con una concepción no-científica de las fases de la Luna (Trundle *et al.*, 2002).
- En lo que respecta a las estaciones del año, otro estudio en EEUU muestra que solo 1 docente en formación sobre 49 encuestados muestra un conocimiento adecuado de la explicación del fenómeno. Entre los demás: la primera CAM para la explicación es la distancia variable Tierra-Sol, como fue señalado antes. Una segunda CAM que se detectó es la inclinación variable del eje de rotación de la Tierra (Atwood y Atwood, 1996). Estudios realizados en Israel indican que este es el tema más difícil (Trumper, 2003).

**¿Qué dicen los programas de estudio sobre lo que deben aprender los docentes?**

Limitándose a la astronomía y a la astrofísica, los programas del gobierno de la Provincia de Buenos Aires incluyen, entre otros, los siguientes temas:

*Docentes para primaria (EGB 1 y 2).*
Ciencias naturales y su enseñanza I: La luz. El sol como fuente de luz. Propagación. Fenómenos de reflexión y refracción de la luz. Movimientos de la Tierra. Las estaciones. Predicción de fenómenos a partir de modelos
Ciencias naturales y su enseñanza II: El hombre en el espacio. Exploración del Sistema Solar. Leyes de la reflexión y refracción de la luz. Descomposición de la luz. Instrumental óptico.
Ciencias naturales y su enseñanza III: Fenómenos ondulatorios. Modelos explicativos en la enseñanza de contenidos científicos.
*Docentes para el EGB-3 y polimodal.*
Ciclo Común-1: Sistema Solar, calendarios. Estaciones del año.
Ciclo Común-2: Evolución de las estrellas. Gravitación universal. Objetos estelares: gigantes rojas, novas, enanas blancas, estrellas de neutrones, agujeros negros.

Profesorado en Física, 3er año: Instrumental astronómico. Espectroscopía. Estructura a gran escala del Universo. Reacciones nucleares y evolución estelar.

Profesorado en Física, 4to año: Cosmología y evolución del Universo: Teorías alternativas. Geometría espacio-temporal.

Por su parte, la Ciudad de Buenos Aires propone:

*Docentes para primaria*

Enseñanza de Ciencias Naturales I y II: Principios de la mecánica - el movimiento de los cuerpos celestes: Los modelos en el campo de la Física y su complejidad. Las leyes de la mecánica y las ideas intuitivas sobre las fuerzas y el movimiento. Cosmovisión acerca de la Tierra y el lugar del hombre en el Universo: modelos geocéntricos y heliocéntricos. El lugar relativo de la observación en función de los puntos de vista del observador.

Nuevamente, en ambos lineamientos curriculares, encontramos algunos de los temas que presentan muy arraigadas ideas alternativas, como ser la explicación de las estaciones del año, y donde las clásicas imágenes de elipses muy elongadas para las trayectorias de los planetas (como la Tierra) pueden inducir a los futuros docentes a aceptar sin mucho cuestionamiento la teoría del alejamiento, o incluso la de la inclinación del eje terrestre, para explicar las diferentes temperaturas o duraciones del día en una u otra estación del año.

**Y los científicos: ¿Qué pueden hacer? ¿Qué deben aprender?**

Como mencionamos al comienzo, no solo los alumnos y los docentes en formación presentaban problemas en temas de astronomía. A los científicos -y entre ellos, a los astrónomos- también les sucede. Aunque en este caso el problema más significativo es el de transmitir sus conocimientos: hacerlo en tal forma que colabore a zanjar la brecha que existe entre el que trabaja en una ciencia, el que la enseña y divulga, y el alumno o, más ampliamente, el público general.

Entre las dificultades de comunicación que se encuentran más a menudo, podemos señalar las concesiones que se deben hacer a veces ante la disyuntiva "comunicación efectiva vs. exactitud" de algunos temas, y la dificultad de encontrar la representación pictórica adecuada de algunos procesos físicos avanzados, entre otras vicisitudes. Problemas similares se presentan a todos los científicos a la hora de colaborar con los profesionales formadores de futuros docentes, pues al trabajar juntos deben compartir un lenguaje común, y no solo eso. Los científicos interesados en la enseñanza de las ciencias deben también conocer los problemas y obstáculos más comunes que existen en la enseñanza-aprendizaje de su ciencia y reflexionar sobre cómo ayudar a resolverlos.

**Investigación: ¿Por qué indagar nociones alternativas en astronomía?**

La extensión y dificultad intrínseca de los temas a ser cubiertos en los programas (en algunos casos más que en otros, por supuesto) nos hacen pensar que las ideas alternativas y otras dificultades de aprendizaje de nociones básicas de la astronomía, ya evidenciadas en diversos estudios internacionales, pueden también darse en nuestras escuelas, tanto en el nivel inicial como en el secundario y en docentes en formación. Evidentemente, sin consideración de estas dificultades en el aprendizaje no puede existir innovación pedagógica. Sin la posibilidad de que la astronomía (por ejemplo) sea bien impartida y aprendida en la escuela primaria, no vemos bien cómo articular un pasaje suave a la secundaria.

La disparidad de niveles culturales de los alumnos y de formación de los docentes no nos permite hacer simplificaciones rápidas, pues en los últimos años la escuela, más que un lugar de nivelación educativa, creemos, se ha mostrado como una barrera de separación de clases sociales.

Luego, no podemos decir que todos los alumnos o que todos los docentes en formación traen estos conceptos no científicos de difícil superación. Sin embargo, es claro que una gran parte de los individuos merecería que se los ayudara a aprender mejor, en el caso de los alumnos, y a enseñar mejor las ciencias, en el caso de los futuros docentes. Se hace necesario entonces indagar sobre ideas alternativas. Es de esperar que, una vez llevado a cabo el trabajo surjan varias nociones alternativas como las ya mencionadas más arriba. Se espera también que se produzca lo que los investigadores llaman acoplamiento de nociones: un conocimiento científico y una noción no-científica aparecen juntas para explicar un dado fenómeno (por ejemplo, para el ciclo día/noche: la rotación más la revolución de la Tierra). Se verá asimismo si existe sedimentación de conocimientos. Esta última terminología indica los casos en que a una vieja (y fuerte) noción equivocada se le agrega la explicación científica, la que, sin embargo, no logra reemplazar a la primera.

En base a los resultados, se verá sin dificultad si la población de los futuros docentes y la de los alumnos testeados presentan o no los mismos problemas de enseñanza-aprendizaje que las poblaciones de otros países (se esperaría que sí).

**Nuestra intervención**

Lo expuesto brevemente en esta presentación es lo que motiva a los autores a iniciar acciones concretas en relación al área de la investigación y educación en astronomía. Para tal fin, nuestra intervención pretende involucrar, entre otros, los siguientes aspectos:

*Primera fase:* Cuestionarios y entrevistas a los docentes y futuros docentes de la enseñanza primaria, para conocer sus representaciones sobre los temas presentados anteriormente. ¿Cómo hacerlo? Aquí no parece necesario innovar demasiado, pues con cuestionarios adecuadamente confeccionados uno puede detectar preconcepciones científicas en un basto rango de temas de astronomía. La innovación entonces sería hacer efectivo este necesario trabajo de campo, como primera medida para detectar deficiencias en el sistema educativo. Cuestionarios completados por futuros docentes antes de que reciban instrucción formal en temas de astronomía, y también luego de completada su formación, permitirán saber si a la unidad El planeta Tierra y el Universo se le da el debido peso en sus estudios y, de no ser así, seleccionar los cambios de metodología a implementar.

Para estos cuestionarios, el método de opción múltiple (*multiple choice*) no es el más adecuado, pues se ha visto que muchas veces, aun sin conocer la verdadera respuesta a una dada pregunta, la sola lectura de la respuesta correcta ayuda al interrogado. Además, al darle libertad para que escriba los justificativos que él/ella crea convenientes uno puede indagar más sobre la lógica que se emplea para seleccionar una u otra respuesta. Esto, evidentemente, ayuda a enriquecer el trabajo del investigador.

*Segunda fase:* Investigaciones sobre las representaciones de los alumnos que actúan como obstáculos para el aprendizaje de contenidos relacionados con el área. Si bien las representaciones estudiadas desde un punto de vista individual pueden presentar rasgos particulares es cierto que, socialmente, reflejan cosmovisiones y modelos. Nuevamente, aquí es de esperarse resultados similares a los encontrados en otros países. El método a utilizar será la elaboración de cuestionarios que permitan explicitar los modelos iniciales que traen los alumnos.

*Tercera fase:* Diseño conjunto entre formadores de docentes e investigadores de materiales didácticos innovadores a fin de generar reales herramientas de apoyo. Estas herramientas deberán tomar como referencia las representaciones de los alumnos y docentes que ya discutimos de manera tal que favorezcan un aprendizaje significativo. Para tal fin, es necesario no sólo capacitar/formar adecuadamente a los futuros docentes en temas de astronomía sino también ofrecer situaciones de enseñanza que expliciten de qué manera hacer frente a estos obstáculos de aprendizaje. Qué actividades seleccionar y cómo orientarlas de manera tal que promuevan la evolución de las concepciones/modelos iniciales de los alumnos a los modelos científicos/escolares deseados.

**Conclusiones**

Se señalaron algunas dificultades de alumnos y alumnas (y niños y niñas en general) en temas de astronomía. Son de esperar inconvenientes similares en el caso de docentes en formación (y hasta en científicos y científicas, que no necesariamente sabemos cómo transmitir nuestros conocimientos). Institutos formadores de docentes, en general, no necesariamente los capacitan adecuadamente en temas de astronomía (¿o sí?). Convendría entonces alentar más el trabajo en talleres específicos, fomentar el contraste de ideas e informaciones, descubrir inconsistencias y anomalías (porque sus modelos iniciales no permiten responder o resolver situaciones problemáticas), intentar la construcción de conocimientos científicos robustos y evitar la sedimentación de conocimientos.
Asimismo, consideramos que los educadores de docentes, e incluso los astrónomos interesados en la educación de su ciencia, deberían prestar especial atención a las concepciones "alternativas" de los futuros docentes en temas de "astro".
Algunas posibles intervenciones a tener en cuenta podrían ser: Científicos y formadores, ¿trabajan juntos?, ¿se les valora ese trabajo? Quizás convendría hacerlo. No se debería descuidar la formación de los docentes, cuando aún son "alumnos". Después cuesta más. ¿Qué aportes hace la didáctica al desarrollo de este campo de investigación? ¿Se ha llevado a cabo un relevamiento en los últimos años sobre ideas previas en astronomía de los (actuales y futuros) docentes (de ciencias)? En este sentido esperamos mediante nuestro trabajo contribuir con el diagnóstico de la problemática de la enseñanza-aprendizaje en temas de astronomía y con la implementación de herramientas pedagógicas innovadoras tendientes a mejorar su enseñanza formal.

**Bibliografía**